\documentclass[11pt]{article}

\usepackage{geometry}                		
\usepackage{graphicx}	
\usepackage[hidelinks]{hyperref}			
\usepackage{amsmath,amssymb}

\newcommand\blfootnote[1]{
    \begingroup
    \renewcommand\thefootnote{}\footnote{#1}
    \addtocounter{footnote}{-1}
    \endgroup
}
\newcommand{\be}{\begin{equation}}
\newcommand{\ee}{\end{equation}}
\newcommand{\ttb}{{T\bar{T}}}
\newcommand{\CE}{{\mathcal{E}}}
\newcommand{\CL}{{\mathcal{L}}}
\newcommand{\CM}{{\mathcal{M}}}
\newcommand{\CN}{{\mathcal{N}}}
\newcommand{\CO}{{\mathcal{O}}}

\newcommand{\CS}{{\mathcal{S}}}

\newcommand{\p}{\partial}
\newcommand{\bp}{\bar{\partial}}
\newcommand{\half}{\frac{1}{2}}

\newcommand{\bz}{\bar{z}}
\newcommand{\bt}{\bar{t}}

\newcommand{\tr}{\mathrm{tr}\,}

\begin{document}
\begin{titlepage}
\flushright RCHEP-24-001\\
\hfill \\
\vspace*{15mm}
\begin{center}
{\Large \bf Holographic Action Principle for $T\bar{T}$-deformation}

\vspace*{15mm}

{\large $\text{Amin Faraji Astaneh}^{a,b}\blfootnote{\href{mailto:faraji@sharif.edu}{email: faraji@sharif.edu}}$}
\vspace*{8mm}

\parbox{ \linewidth}{\begin{center}$^a$Department of Physics, Sharif University of Technology,\\
P.O.Box 11155-9161, Tehran, Iran\\ \vspace{0.5cm} 
$^b$Research Center for High Energy Physics\\
Department of Physics, Sharif University of Technology,\\
P.O.Box 11155-9161, Tehran, Iran
\end{center}}\\

\vspace*{0.7cm}

\end{center}
\begin{abstract}
We explore the action principle for the holographic $\ttb$-deformation. We develop a scheme in which one can holographically reproduce the action of the Liouville theory deformed by $\ttb$-insertion. This scheme necessitates considering the bending energy of the finite cut-off surface. Following our proposal, one observes a perfect match between the actions of deformed theory on the field theory and gravity sides.
\end{abstract}

\end{titlepage}

\newpage
\section{Introduction}
The problem of $\ttb$-deformation of field theories has sparked significant interest in a wide range of fields, ranging from statistical physics to high-energy physics. The unique characteristic of this deformation is that it is irrelevant but still integrable, which is why it has garnered special attention \cite{Zamolodchikov:2004ce}-\cite{Cavaglia:2016oda}. 

This deformation is implemented by inserting a composite operator, $\CO_{T\bar{T}}$ constructed from the determinant of the energy-momentum tensor. Denoting the action of the deformed theory as
\be
S_\lambda=\int d^2x\sqrt{h}\, \CL_\lambda\, ,
\ee
where $\lambda$ is the parameter of deformation, the flow equation that describes the deformation reads
\be\label{flow}
\p_\lambda S_\lambda=\int\sqrt{h}\, \CO_\ttb\, .
\ee
The integrability means that while this flow takes us away from the original theory, certain quantities can still be obtained in terms of the corresponding values in the original theory. The energy spectrum of the deformed theory is one of those things.

The dimensionless energy of the deformed theory computed on a cylinder of circumference $2\pi$ lies in the following spectrum (see {\textit{e.g.}} \cite{Smirnov:2016lqw} and \cite{Cavaglia:2016oda})
\be\label{energy CFT}
\CE_n=\frac{2\pi^2}{\lambda}\left(1-\sqrt{1-\frac{2\lambda}{\pi}M_n+\frac{\lambda^2}{\pi^2}J_n^2}\,\right)\ , \ 
\begin{cases}
M_n=\Delta_n+\bar{\Delta}_n-\frac{c}{12}\, ,\\
J_n=\Delta_n+\bar{\Delta}_n\, ,
\end{cases}
\ee
where $c$ is the central charge and $(\Delta_n,\bar{\Delta}_n)$ are the conformal dimensions of the state of the original CFT.

Interestingly, there is an intuitive holographic picture in correspondence with this special kind of deformation. This holographic prescription, which is known as the \emph{holography at finite cut-off}, takes its main ideas from the holographic renormalization group \cite{V}, see also \cite{Guica}-\cite{Tian:2024vln} and \cite{Asrat:2017tzd}-\cite{Giveon:2017nie} for a different proposal in the context of UV/IR interpolating backgrounds in string theory. The proposal suggests removing the asymptotic region of AdS$_3$ and looking at the field theory which now lives on a finite radius cut-off (neither at infinity nor at zero, depending on the choice of the coordinate). The proposal is that the theory living on this finite cut-off coincides with the deformed theory with $\ttb$-insertion. The duality implies that the radius of this finite cut-off, $r_c$ will have a simple relation with the deformation parameter, $\lambda$.
A simple way to check this prescription is to calculate the energy spectrum, holographically.  
One can reproduce \eqref{energy CFT}, computing the quasi-local energy of a BTZ black hole (with mass $M$ and angular momentum $J$) at a finite cut-off $r_c$, \cite{V}. This calculation yields the proportionality $\lambda\propto r_c^{-2}$, 
as previosly assured. The proportionality coefficient will be determined in terms of the Newton constant or equivalently the central charge, which will be explicitly determined later.

This proposal has been verified through the calculation of other quantities such as thermodynamic quantities, entanglement entropy (\cite{Donnelly:2018bef}-\cite{FarajiAstaneh:2022qck}), correlation functions (\cite{Hartman:2018tkw}, and see \cite{Cardy:2019qao} for a CFT calculation), and so on. See also \cite{Bhattacharyya:2023gvg}-\cite{He:2024xbi} for recent developments. However, the accurate description of the duality through the matching of partition functions of field theory and gravity remains to be explored. In this paper, we will try to explore such an equivalence in the context of the correspondence between a Liouville theory and its gravitational dual theory.

The paper is organized as follows: 

In section 2 we examine the Liouville theory and its deformation by $\ttb$-insertion. Section 3 focuses on the holographic reproduction of the deformed action, where we present our holographic proposal. Finally, section 4 concludes this study.

\section{Liouville theory and its $\ttb$-deformation}
The simplest model that is nontrivial enough for us to test our proposal against is the Liouville model, for earlier treatments of deformation in scalar theory, see \cite{Cavaglia:2016oda}, \cite{Bonelli:2018kik} and \cite{Leoni:2020rof}.

In this paper, our focus will be on the Liouville field theory with the following action\footnote{Here, we intentionally employed subscript $0$ for the Lagrangian and action, because we intend to deform the theory by $\ttb$ insertion and obtain the higher-order Lagrangian in subsequent advancements.}

\be
S_0=\frac{1}{2\kappa}\int d^2x\sqrt{h}\CL_0=\frac{1}{2\kappa}\int d^2x\sqrt{h}(\half h^{ij}\p_i\phi\p_j\phi+\phi R_h)\, .
\ee
We will suitably set the prefactor $\kappa$ later on.

To simplify the writing of the formulas, we define the two tensors $r_{ij}$ and $s_{ij}$ as follows

\be
r_{ij}=\p_i\p_j\phi\ , \ s_{ij}=\p_i\phi\p_j\phi\, ,
\ee
so 
\be
\CL_0=\frac{s}{2}+\phi R_h\, .
\ee
We should emphasize that the metric $h_{ij}$ is auxiliary and will choose properly to give us the action in its appropriate form. We will construct all contractions of $r_{ij}$ and $s_{ij}$ using this metric, for instance, $s=h^{ij}s_{ij}$.
The equation of motion fixes the curvature tensor as $R_h=r$.

For this theory, the energy-momentum tensor with definition
\be
T^{0}_{ij}=-\frac{2}{\sqrt{h}}\frac{\delta S_0}{\delta h^{ij}}=h_{ij}\CL_0-2\frac{\delta\CL_0}{\delta h^{ij}}\, ,
\ee
would be determined as
\be
T^{0}_{ij}=2(r_{ij}-h_{ij}r)-(s_{ij}-\half h_{ij}s)\, .
\ee
In continuation, we work in conformal gauge with the Liouville field as the rescaling parameter that sets the stage for our future objectives. Covering the two dimensional base manifold, $\CM$ with  the complex coordinates $(z,\bz)$, the metric reads
\be
ds^2_\CM=h_{ij}dx^idx^j=e^{-\phi(z,\bz)}dz d\bz\, .
\ee
In this gauge
\be
R_h=r=4e^\phi\p\bp\phi\ , \ s=4e^\phi\p\phi\bp\phi\, ,
\ee
and defining
\be
\begin{cases}
t=r_{zz}-\half s_{zz}=(\p\phi)^2-\half \p^2\phi \\
\bt=r_{\bz\bz}-\half s_{\bz\bz}=(\bp\phi)^2-\half \bp^2\phi \, ,
\end{cases}
\ee
one explicitly finds that
\be
T^0_{zz}=t \ , \ T^0_{\bz\bz}=\bt \ , \ T^0_{z\bz}=T^0_{z\bz}=-\frac{1}{4}re^{-\phi}\, .
\ee
We are now fully equipped to investigate the $\ttb$-deformation of the theory. Let us start with the basic definition of the deformation operator, $\CO_\ttb$ which is a composite operator defined as
\be
\CO_\ttb=\half\epsilon^{ik}\epsilon^{j\ell}T^\lambda_{ij}T^\lambda_{k\ell}\, ,
\ee
where
\be
T^\lambda_{ij}=-\frac{2}{\sqrt{h}}\frac{\delta S_\lambda}{\delta h^{ij}}=h_{ij}\CL_\lambda-2\frac{\delta\CL_\lambda}{\delta h^{ij}}\, ,
\ee
is the energy momentum tensor of deformed theory.
Using this definition one explicitly finds \cite{Bonelli:2018kik}
\be\label{ttb def}
\CO_\ttb=\CL_\lambda^2-2\CL_\lambda\, h^{ij}\frac{\delta\CL_\lambda}{\delta h^{ij}}+2\epsilon^{ik}\epsilon^{j\ell}\frac{\delta\CL_\lambda}{\delta h^{ij}}\frac{\delta\CL_\lambda}{\delta h^{k\ell}}\, .
\ee
Expanding the deformed Lagrangian in powers of parameter of deformation as
\be
\CL_\lambda=\sum_{k=0}\lambda^k\CL_k\, ,
\ee
one can perturbatively track the deformation. To the first order, one must substitute $\CL_0$ in the right-hand side of \eqref{ttb def}, then solve \eqref{flow}. Doing so one finds
\be
\CL_1=\frac{1}{2\kappa^2}\left[r^{ij}(s_{ij}-r_{ij})+r^2-\half rs-\frac{1}{8}s^2\right]+\frac{1}{\kappa^2}\left[\phi(r^{ij}-\half s^{ij})-\half\phi^2R_h^{ij}\right]G^h_{ij}\, ,
\ee
where $G^h_{ij}$ is the Einstein tensor on the two-dimensional base manifold. since the Einstein tensor vanishes identically in two dimensions, $G^h_{ij}=0$ and the second bracket drops out.

In conformal gauge, one explicitly finds
\be
\CL_1=\frac{1}{2\kappa^2}(\frac{r^2}{4}e^{-\phi}-4e^{\phi}t\bar{t})\, ,
\ee
and thus
\be
S_1=\frac{1}{4\kappa^2}\int dzd\bz\, (\frac{r^2}{4}e^{-2\phi}-4t\bar{t})\, ,
\ee
whereby $S_1$ we mean the deformed action of the Liouville theory at the first order of perturbation in terms of the parameter of deformation. Therefore, up to the leading order of deformation
\be\label{SL}
S_L=\frac{1}{4\kappa}\int dzd\bz\, e^{-\phi}(\frac{s}{2}+\phi r)+\frac{\lambda}{4\kappa^2}\int dzd\bz\left(\frac{r^2}{4}e^{-2\phi}-4t\bt\right)\, .
\ee
This is the main result on the field theory side that we seek to replicate holographically.
\section{Holography}
A rigorous statement of the holographic principle, which is known as the AdS/CFT correspondence, proposes an equivalence between the partition functions of a gravitational theory in AdS spacetime and the dual conformal field theory living on its boundary \cite{Maldacena:1997re}-\cite{Aharony:1999ti}.

In the context of Liouville theory, holographic investigations have a rich history; for instance, one can refer to the attempts presented in \cite{Coussaert:1995zp} to \cite{Li:2019mwb}. In this study, we employ the machinery of holographic reconstruction of spacetime to construct a three-dimensional bulk space in which evaluating the gravitational on-shell action resembles the Liouville theory in the near-boundary limit.

Doing so, we utilize the fact that the Liouville action appears as a result of Weyl rescaling of the metric on the asymptotic boundary of AdS$_3$. The Liouville field coincides with the field of the Weyl rescaling. Referring to AdS/CFT correspondence one needs to evaluate the gravitational action in the bulk with a non-constant radius boundary whose profile is determined by the Liouville field, to reproduce the Liouville action on it, see \cite{Nguyen:2021pdz} for a related study.

To explicitly demonstrate this,
let us run the machinery of holographic reconstruction of spacetime in Fefferman-Graham coordinates, \cite{2007arXiv0710.0919F}, \cite{Skenderis:1999nb} and \cite{deHaro:2000vlm}.
\be
ds^2_\CN=\frac{d\rho^2}{4\rho^2}+\frac{1}{\rho}g_{ij}(x,\rho)dx^idx^j\, , \, g_{ij}(x,\rho)=g^{(0)}_{ij}+\rho g^{(1)}_{ij}+\rho^2 g^{(2)}_{ij}\, .
\ee
The asymptotic metric $g^{(0)}_{ij}$ is naturally a conformally flat metric in which we deliberately identify the conformal field with the Liouville field.
In this way, the Liouville field contributes to the reconstruction of the bulk spacetime.
Specifically, the Einstein equation fixes the coefficient $g^{(1)}_{ij}$ (and then indirectly $g^{(2)}_{ij}$) up to a conserved symmetric tensor whose trace is proportional to the Ricci scalar. This tensor serves as the energy-momentum tensor of the Liouville field, establishing the pathway through which the Liouville field shapes the metric of the bulk space.
More accurately, 
\be
g^{(1)}_{ij}=-\half\left(g^{(0)}_{ij}R^{(0)}+\kappa T^{(0)}_{ij}\right)\, , \, g^{(2)}_{ij}=\frac{1}{4}g^{(0)k\ell}g^{(1)}_{ik}g^{(1)}_{j\ell}\, .
\ee
So adopting the complex coordinate system and setting $g^{(0)}_{ij}dx^idx^j=e^{\phi(z,\bz)}dzd\bz$,
one explicitly arrives at
\begin{align}
ds^2_\CN&=\frac{d\rho^2}{4\rho^2}+\frac{1}{2\rho}\left[\left(t\rho+\frac{rt}{8}e^{-2\phi}\rho^2\right)dz^2+\left(\bt\rho+\frac{r\bt}{8}e^{-2\phi}\rho^2\right)d\bz^2\right]\\
&+\frac{1}{\rho}\left[e^\phi+\frac{r}{4}e^{-\phi}\rho+\frac{1}{4}e^{-\phi}\left(t\bt+\frac{r^2}{16}e^{-2\phi}\right)\rho^2\right] dzd\bz\, .
\end{align}
A key aspect of this holographic completion is that performing the following change of coordinates brings us to the standard Poincaré coordinates, \cite{Krasnov:2001cu} and \cite{Hung:2011nu}
\be
r=e^{-\frac{\phi}{2}}(\rho^{-\frac{1}{2}}e^\phi+\frac{s}{16}\rho^{\frac{1}{2}}e^{-\phi})\, ,
\ee
and
\be
(z,\bz)\rightarrow (z,\bz)+\frac{\rho}{2}\frac{1}{e^\phi+\frac{s}{16}\rho e^{-\phi}}(\bp\phi,\p\phi)\, .
\ee
The presence of the Liouville field on the right-hand side of the first relation clarifies why we assert that the evaluation of the gravitational action in a bulk space with a non-constant radius cutoff may yield the action of the Liouville theory. 
We are ready now to evaluate the gravitational action. This evaluation is presented through the following sub-sections.
\subsection{Einstein-Hilbert action}
The Einstein-Hilbert action reads
\be
S_{EH}=\frac{1}{2\kappa'}\int d^3X\sqrt{G}(R_G-2\Lambda)=-\frac{2}{\kappa'}\int dzd\bz\int d\rho\sqrt{G}\, ,
\ee
where $\kappa'=8\pi G_N$ and the integral over $\rho$ is taken from $\rho_c e^{\phi}$ as mentioned before. A simple calculation yields that up the linear order of the finite cut-off radius
\be
S_{EH}=-\frac{V_2}{2\kappa'}\frac{1}{\rho_c}+\frac{1}{8\kappa'}\int dzd\bz\, e^{-\phi}(\phi r+r\log\rho_c)+\frac{1}{8\kappa'}\int dzd\bz\left(\frac{r^2}{16}e^{-2\phi}-t\bt\right)\rho_c\, ,
\ee
where $V_2$ is the volume of the two-dimensional base manifold, $\CM$\footnote{We only look at the asymptotic terms and hence disregard the IR divergences for the moment.}.\\
\subsection{Gibbons-Hawking action and the counterterm}
Locating the asymptotic boundary at $\rho_ce^\phi$, the following metric will be induced on that
\be
ds^2_{ind}=\gamma_{ij}dx^idx^j \ , \ \gamma_{ij}=\frac{1}{4}s_{ij}+\frac{e^{-\phi}}{\rho_c}g_{ij}(x,\rho_ce^\phi)\, .
\ee
The Gibbons-Hawking action reads then
\be
S_{GH}=-\frac{1}{\kappa'}\int dzd\bz\sqrt{\gamma}K\, ,
\ee
where $K$ is the trace of the extrinsic curvature of the finite cut-off surface. This boundary term should be accompanied by the following simple counterterm, as is well known.

\be
S_{c.t.}=-\frac{1}{\kappa'}\int dzd\bz\sqrt{\gamma}\, ,
\ee
 Up the linear order of the finite cut of radius, these boundary actions read
 \be
\begin{split}
&S_{GH}=\frac{V_2}{\kappa'}\frac{1}{\rho_c}+\frac{1}{8\kappa'}\int dzd\bz\, e^{-\phi}(s+2r)\\
&+\frac{1}{64\kappa'}\int dzd\bz e^{-2\phi}\left[\frac{1}{4}\tr(s^2)-2\tr(rs)+2\tr(r^2)-2r^2+rs\right]\rho_c\, ,
\end{split}
\ee
and
\be
\begin{split}
&S_{c.t.}=-\frac{V_2}{2\kappa'}\frac{1}{\rho_c}-\frac{1}{16\kappa'}\int dzd\bz\, e^{-\phi}(s+2r)\\
&+\frac{1}{128\kappa'}\int dzd\bz e^{-2\phi}\left[-\frac{1}{4}\tr(s^2)+2\tr(rs)+2\tr(r^2)-2r^2-rs\right]\rho_c\, ,
\end{split}
\ee
which finally gives
\be\label{GHct}
\begin{split}
&S_{GH}+S_{c.t.}=\frac{V_2}{2\kappa'}\frac{1}{\rho_c}+\frac{1}{16\kappa'}\int dzd\bz\, e^{-\phi}(s+2r)\\
&+\frac{1}{128\kappa'}\int dzd\bz e^{-2\phi}\left[\frac{1}{4}\tr(s^2)-2\tr(rs)+6\tr(r^2)-6r^2+rs\right]\rho_c\, ,
\end{split}
\ee
where $\tr(s^2)=h^{ik}h^{j\ell}s_{ij}s_{k\ell}$, and so on. We would like to highlight the sub-leading term of the boundary terms, as it will interestingly cancel out in the subsequent steps of the proposal, as follows.
\subsection{Bending (Willmore) energy of the finite cut-off surface}
Upon revisiting the AdS/CFT correspondence one knows that moving along the radial coordinate of the bulk corresponds to moving along the RG trajectory. In this sense, the different radii in the bulk are energetically distinct. Therefore, bending a surface at a finite cutoff radius requires different energy than bending it at infinity (or at zero radius).

The bending energy of surface $\CS$ will be determined by the Willmore function which is the integral of the square of the trace of the extrinsic curvature on $\CS$, \cite{W} and \cite{Astaneh:2014uba} 
\be
W=\frac{1}{2}\int_{\CS}K^2\, .
\ee
Therefore referring to the holography at finite cut-off, one needs to subtract the following term from the total gravitational action
\be
S_{W}=\frac{1}{2\kappa'}\left(W\vert_{\rho=\rho_ce^{\phi}}-W\vert_{\rho=0}\right)\, .
\ee
Interestingly enough
\be
S_W=\frac{1}{128\kappa'}\int dzd\bz e^{-2\phi}\left[\frac{1}{4}\tr(s^2)-2\tr(rs)+6\tr(r^2)-6r^2+rs\right]\rho_c\, ,
\ee
and this is exactly what we get from the Gibbons-Hawking action plus the counterterm at $\CO(\rho_c^1)$ order, \eqref{GHct}.

Putting things together and dropping the total derivative terms, one ultimately gets
\be
\begin{split}
S_{grav}&=S_{EH}+S_{GH}+S_{c.t.}-S_W\\
&=\frac{1}{8\kappa'}\int dzd\bz\, e^{-\phi}(\frac{s}{2}+\phi r)+\frac{1}{32\kappa' r_c^2}\int dzd\bz\left(\frac{r^2}{4}e^{-2\phi}-4t\bt\right)\, .
\end{split}
\ee
Comparing this result with \eqref{SL}, it becomes evident that we have reproduced the deformed Liouville action holographically. One simply needs to take the final steps and match the prefactors in these two structurally equivalent actions. This is what we will address in the next sub-section.
\subsection{Matching the actions via AdS/CFT correspondance}
As mentioned above, the gravitational action, when taking the bending energy of the boundary into account, fully replicates the deformed Liouville action. By demanding the equality of the prefactors, one finds
\be
\text{At the order of $\lambda^0$:}\ \kappa=2\kappa'=16\pi G_N=\frac{24\pi}{c}\, ,
\ee
where in the last step we have employed the Brown-Hennaux relation, $\frac{1}{2G_N}=\frac{c}{3}$ \cite{BH}.

Equating the actions at the next order gives\footnote{One should note that our definition of $\CO_\ttb$ differs from \cite{V} by some numerical factors, this causes a numerically different relation between $\lambda$ and $r_c$.}
\be
\text{At the order of $\lambda^1$:}\ \lambda=\frac{\kappa'}{4r_c^2}=\frac{2\pi G_N}{r_c^2}=\frac{3\pi}{cr_c^2}\, .
\ee
The obtained pre-factors are consistent with the expected values and have been presented in the literature using some alternative calculations. Presenting these results we conclude our calculations at this point.
\section{Conclusion}
In this paper, we have developed a gravitational framework that enables the holographic reproduction of the action of the field theory deformed by $\ttb$-insertion. The significance of this work lies in the fact that the gravitational action in the usual holographic reconstruction of spacetime is scheme dependent, resulting in finite undetermined terms. However, in pursuit of our new objective which is investigating the deformed theory within the holographic framework of the finite cut-off proposal, these terms find crucial roles. Therefore one needs to design a precise and unambiguous scheme to appropriately produce the action of the deformed field theory on the asymptotic boundary.

A very crucial point in designing this scheme is that bending the boundary in a region with a finite radius in bulk space introduces the Willmore energy on the surface, and we must incorporate this term into the total gravitational action. Introducing this bending term appropriately is crucial for accurately reproducing the action of the field theory. Adding this term provides us with a very simple recipe for the regularization we need to impose to get the action of deformed theory. In fact, in this scheme, all that we need is to calculate the contribution of the volume part of the gravitational action at the leading order of deformation. By carefully considering the Willmore energy, all unintended boundary terms at the leading order of deformation cancel each other out, leaving us with the deformed field theory action. To the best of our knowledge, this procedure is entirely novel, and we have tested it for the Liouville field theory, which is simple yet rich enough and nontrivial.

Given that knowing the action and partition function is the first step in generic calculations in field theory and holography, the results we have obtained and the scheme we have fixed can be utilized in many other calculations. For example, one could explore the modified gravitational actions in this scheme and examine their dual deformed theories. Additionally, further exploration of certain quantum information quantities such as the quantum complexity, which is closely bound up with evaluation of the gravitational action, could be among the next objectives of this study. The next step of these calculations could be a thorough investigation of deformed theories that, in their original form, exhibit specific symmetries, such as non-relativistic theories. We set aside these studies for future works.


\begin{thebibliography}{100}

\bibitem{Zamolodchikov:2004ce}
A.~B.~Zamolodchikov,
``Expectation value of composite field T anti-T in two-dimensional quantum field theory,''
[arXiv:hep-th/0401146 [hep-th]].

\bibitem{Smirnov:2016lqw}
F.~A.~Smirnov and A.~B.~Zamolodchikov,
``On space of integrable quantum field theories,''
Nucl. Phys. B \textbf{915}, 363-383 (2017)
doi:10.1016/j.nuclphysb.2016.12.014
[arXiv:1608.05499 [hep-th]].

\bibitem{Cavaglia:2016oda}
A.~Cavagli\`a, S.~Negro, I.~M.~Sz\'ecs\'enyi and R.~Tateo,
``$T \bar{T}$-deformed 2D Quantum Field Theories,''
JHEP \textbf{10}, 112 (2016)
doi:10.1007/JHEP10(2016)112
[arXiv:1608.05534 [hep-th]].


		
		\bibitem{V}
		L.~McGough, M.~Mezei and H.~Verlinde,
		``Moving the CFT into the bulk with $ T\overline{T} $,''
		JHEP \textbf{04}, 010 (2018)
		doi:10.1007/JHEP04(2018)010
		[arXiv:1611.03470 [hep-th]].
		
\bibitem{Guica}
M.~Guica and R.~Monten,
``$T\bar T$ and the mirage of a bulk cutoff,''
SciPost Phys. \textbf{10}, no.2, 024 (2021)
doi:10.21468/SciPostPhys.10.2.024
[arXiv:1906.11251 [hep-th]].

\bibitem{Kraus:2018xrn}
P.~Kraus, J.~Liu and D.~Marolf,
``Cutoff AdS$_{3}$ versus the $ T\overline{T} $ deformation,''
JHEP \textbf{07}, 027 (2018)
doi:10.1007/JHEP07(2018)027
[arXiv:1801.02714 [hep-th]].

		\bibitem{Taylor}
		M.~Taylor,
		``TT deformations in general dimensions,''
		[arXiv:1805.10287 [hep-th]].
		
\bibitem{Tian:2024vln}
J.~Tian, T.~Lai and F.~Omidi,
``Modular transformations of on-shell actions of (root-)$\text{T}\overline{\text{T}}$ deformed holographic CFTs,''
[arXiv:2404.16354 [hep-th]].
		
		
		\bibitem{Asrat:2017tzd}
		M.~Asrat, A.~Giveon, N.~Itzhaki and D.~Kutasov,
		``Holography Beyond AdS,''
		Nucl. Phys. B \textbf{932}, 241-253 (2018)
		doi:10.1016/j.nuclphysb.2018.05.005
		[arXiv:1711.02690 [hep-th]].

		
		\bibitem{Giveon:2017myj}
		A.~Giveon, N.~Itzhaki and D.~Kutasov,
		``A solvable irrelevant deformation of AdS$_{3}$/CFT$_{2}$,''
		JHEP \textbf{12}, 155 (2017)
		doi:10.1007/JHEP12(2017)155
		[arXiv:1707.05800 [hep-th]].
		
		\bibitem{Giveon:2017nie}
		A.~Giveon, N.~Itzhaki and D.~Kutasov,
		``$ \mathrm{T}\overline{\mathrm{T}} $ and LST,''
		JHEP \textbf{07}, 122 (2017)
		doi:10.1007/JHEP07(2017)122
		[arXiv:1701.05576 [hep-th]].
		
		
		

\bibitem{Bonelli:2018kik}
G.~Bonelli, N.~Doroud and M.~Zhu,
``$T \bar{T}$-deformations in closed form,''
JHEP \textbf{06}, 149 (2018)
doi:10.1007/JHEP06(2018)149
[arXiv:1804.10967 [hep-th]].

\bibitem{Leoni:2020rof}
M.~Leoni,
``$ T\overline{T} $ deformation of classical Liouville field theory,''
JHEP \textbf{07}, no.07, 230 (2020)
doi:10.1007/JHEP07(2020)230
[arXiv:2005.08906 [hep-th]].

\bibitem{Donnelly:2018bef}
W.~Donnelly and V.~Shyam,
``Entanglement entropy and $T \overline{T}$ deformation,''
Phys. Rev. Lett. \textbf{121}, no.13, 131602 (2018)
doi:10.1103/PhysRevLett.121.131602
[arXiv:1806.07444 [hep-th]].

\bibitem{Park:2018snf}
C.~Park,
``Holographic Entanglement Entropy in Cutoff AdS,''
Int. J. Mod. Phys. A \textbf{33}, no.36, 1850226 (2019)
doi:10.1142/S0217751X18502263
[arXiv:1812.00545 [hep-th]].

\bibitem{Chen:2018eqk}
B.~Chen, L.~Chen and P.~X.~Hao,
``Entanglement entropy in $T\overline{T}$-deformed CFT,''
Phys. Rev. D \textbf{98}, no.8, 086025 (2018)
doi:10.1103/PhysRevD.98.086025
[arXiv:1807.08293 [hep-th]].

\bibitem{Banerjee:2019ewu}
A.~Banerjee, A.~Bhattacharyya and S.~Chakraborty,
``Entanglement Entropy for $TT$ deformed CFT in general dimensions,''
Nucl. Phys. B \textbf{948}, 114775 (2019)
doi:10.1016/j.nuclphysb.2019.114775
[arXiv:1904.00716 [hep-th]].

\bibitem{Grieninger:2019zts}
S.~Grieninger,
``Entanglement entropy and $ T\overline{T} $ deformations beyond antipodal points from holography,''
JHEP \textbf{11}, 171 (2019)
doi:10.1007/JHEP11(2019)171
[arXiv:1908.10372 [hep-th]].

\bibitem{Allameh:2021moy}
K.~Allameh, A.~F.~Astaneh and A.~Hassanzadeh,
``Aspects of holographic entanglement entropy for TT\textasciimacron{}-deformed CFTs,''
Phys. Lett. B \textbf{826}, 136914 (2022)
doi:10.1016/j.physletb.2022.136914
[arXiv:2111.11338 [hep-th]].

\bibitem{FarajiAstaneh:2022qck}
A.~Faraji Astaneh and K.~Allameh,
``Energy of decomposition and entanglement thermodynamics for T2-deformation,''
Phys. Lett. B \textbf{839}, 137772 (2023)
doi:10.1016/j.physletb.2023.137772
[arXiv:2212.02816 [hep-th]].

\bibitem{Hartman:2018tkw}
T.~Hartman, J.~Kruthoff, E.~Shaghoulian and A.~Tajdini,
``Holography at finite cutoff with a $T^2$ deformation,''
JHEP \textbf{03}, 004 (2019)
doi:10.1007/JHEP03(2019)004
[arXiv:1807.11401 [hep-th]].

\bibitem{Cardy:2019qao}
J.~Cardy,
``$T\bar T$ deformation of correlation functions,''
JHEP \textbf{12}, 160 (2019)
doi:10.1007/JHEP12(2019)160
[arXiv:1907.03394 [hep-th]].

\bibitem{Bhattacharyya:2023gvg}
A.~Bhattacharyya, S.~Ghosh and S.~Pal,
``Aspects of $T\bar{T}+J\bar{T }$ deformed 2D topological gravity : from partition function to late-time SFF,''
[arXiv:2309.16658 [hep-th]].

\bibitem{Babaei-Aghbolagh:2022uij}
H.~Babaei-Aghbolagh, K.~B.~Velni, D.~M.~Yekta and H.~Mohammadzadeh,
``Emergence of non-linear electrodynamic theories from TT\textasciimacron{}-like deformations,''
Phys. Lett. B \textbf{829}, 137079 (2022)
doi:10.1016/j.physletb.2022.137079
[arXiv:2202.11156 [hep-th]].

\bibitem{Babaei-Aghbolagh:2022leo}
H.~Babaei-Aghbolagh, K.~Babaei Velni, D.~Mahdavian Yekta and H.~Mohammadzadeh,
``Marginal TT\textasciimacron{}-like deformation and modified Maxwell theories in two dimensions,''
Phys. Rev. D \textbf{106}, no.8, 086022 (2022)
doi:10.1103/PhysRevD.106.086022
[arXiv:2206.12677 [hep-th]].

\bibitem{Babaei-Aghbolagh:2024hti}
H.~Babaei-Aghbolagh, S.~He, T.~Morone, H.~Ouyang and R.~Tateo,
``Geometric formulation of generalized root-$T\bar{T}$ deformations,''
[arXiv:2405.03465 [hep-th]].

\bibitem{He:2023hoj}
S.~He, Y.~Li, Y.~Z.~Li and Y.~Zhang,
``Holographic torus correlators of stress tensor in AdS$_{3}$/CFT$_{2}$,''
JHEP \textbf{06}, 116 (2023)
doi:10.1007/JHEP06(2023)116
[arXiv:2303.13280 [hep-th]].

\bibitem{He:2023knl}
S.~He, Y.~Z.~Li and Y.~Zhang,
``Holographic torus correlators in AdS$_{3}$ gravity coupled to scalar field,''
JHEP \textbf{05}, 254 (2024)
doi:10.1007/JHEP05(2024)254
[arXiv:2311.09636 [hep-th]].

\bibitem{He:2024fdm}
S.~He, Y.~Li, Y.~Z.~Li and Y.~Zhang,
``Note on holographic torus stress tensor correlators in $AdS_3$ gravity,''
[arXiv:2405.01255 [hep-th]].

\bibitem{He:2024xbi}
S.~He, Y.~z.~Li and Y.~Xie,
``Holographic stress tensor correlators on higher genus Riemann surfaces,''
[arXiv:2406.04042 [hep-th]].

\bibitem{Maldacena:1997re}
J.~M.~Maldacena,
``The Large N limit of superconformal field theories and supergravity,''
Adv. Theor. Math. Phys. \textbf{2}, 231-252 (1998)
doi:10.4310/ATMP.1998.v2.n2.a1
[arXiv:hep-th/9711200 [hep-th]].

\bibitem{Witten:1998qj}
E.~Witten,
``Anti-de Sitter space and holography,''
Adv. Theor. Math. Phys. \textbf{2}, 253-291 (1998)
doi:10.4310/ATMP.1998.v2.n2.a2
[arXiv:hep-th/9802150 [hep-th]].

\bibitem{Aharony:1999ti}
O.~Aharony, S.~S.~Gubser, J.~M.~Maldacena, H.~Ooguri and Y.~Oz,
``Large N field theories, string theory and gravity,''
Phys. Rept. \textbf{323}, 183-386 (2000)
doi:10.1016/S0370-1573(99)00083-6
[arXiv:hep-th/9905111 [hep-th]].

\bibitem{Coussaert:1995zp}
O.~Coussaert, M.~Henneaux and P.~van Driel,
``The Asymptotic dynamics of three-dimensional Einstein gravity with a negative cosmological constant,''
Class. Quant. Grav. \textbf{12}, 2961-2966 (1995)
doi:10.1088/0264-9381/12/12/012
[arXiv:gr-qc/9506019 [gr-qc]].

\bibitem{Verlinde:1989ua}
H.~L.~Verlinde,
``Conformal Field Theory, 2-$D$ Quantum Gravity and Quantization of Teichmuller Space,''
Nucl. Phys. B \textbf{337}, 652-680 (1990)
doi:10.1016/0550-3213(90)90510-K

\bibitem{Krasnov:2000ia}
K.~Krasnov,
``3-D gravity, point particles and Liouville theory,''
Class. Quant. Grav. \textbf{18}, 1291-1304 (2001)
doi:10.1088/0264-9381/18/7/311
[arXiv:hep-th/0008253 [hep-th]].

\bibitem{Krasnov:2002rn}
K.~Krasnov,
``Lambda less than 0 quantum gravity in (2+1)-dimensions. 2. Black hole creation by point particles,''
Class. Quant. Grav. \textbf{19}, 3999-4028 (2002)
doi:10.1088/0264-9381/19/15/309
[arXiv:hep-th/0202117 [hep-th]].

\bibitem{Li:2019mwb}
S.~Li, N.~Toumbas and J.~Troost,
``Liouville Quantum Gravity,''
Nucl. Phys. B \textbf{952}, 114913 (2020)
doi:10.1016/j.nuclphysb.2019.114913
[arXiv:1903.06501 [hep-th]].

\bibitem{Nguyen:2021pdz}
K.~Nguyen,
``Holographic boundary actions in AdS$_{3}$/CFT$_{2}$ revisited,''
JHEP \textbf{10}, 218 (2021)
doi:10.1007/JHEP10(2021)218
[arXiv:2108.01095 [hep-th]].

\bibitem{2007arXiv0710.0919F}Fefferman, C. \& Graham, C. The ambient metric. {\em ArXiv E-prints}. pp. earXiv:0710.0919 (2007,10)

\bibitem{Skenderis:1999nb}
K.~Skenderis and S.~N.~Solodukhin,
``Quantum effective action from the AdS/CFT correspondence,''
Phys. Lett. B \textbf{472}, 316-322 (2000)
doi:10.1016/S0370-2693(99)01467-7
[arXiv:hep-th/9910023 [hep-th]].

\bibitem{deHaro:2000vlm}
S.~de Haro, S.~N.~Solodukhin and K.~Skenderis,
``Holographic reconstruction of space-time and renormalization in the AdS / CFT correspondence,''
Commun. Math. Phys. \textbf{217}, 595-622 (2001)
doi:10.1007/s002200100381
[arXiv:hep-th/0002230 [hep-th]].

\bibitem{2007arXiv0710.0919F}Fefferman, C. \& Graham, C. The ambient metric. {\em ArXiv E-prints}. pp. earXiv:0710.0919 (2007,10)

\bibitem{Krasnov:2001cu}
K.~Krasnov,
``On holomorphic factorization in asymptotically AdS 3-D gravity,''
Class. Quant. Grav. \textbf{20}, 4015-4042 (2003)
doi:10.1088/0264-9381/20/18/311
[arXiv:hep-th/0109198 [hep-th]].

\bibitem{Hung:2011nu}
L.~Y.~Hung, R.~C.~Myers, M.~Smolkin and A.~Yale,
``Holographic Calculations of Renyi Entropy,''
JHEP \textbf{12}, 047 (2011)
doi:10.1007/JHEP12(2011)047
[arXiv:1110.1084 [hep-th]].

\bibitem{W}
T. J. Willmore, 
``Note on Embedded Surfaces,''
An. Sti. Univ. Al. I. Cuza Iasi, N. Ser., Sect.
Ia 11B, 493 (1965).

\bibitem{Astaneh:2014uba}
A.~F.~Astaneh, G.~Gibbons and S.~N.~Solodukhin,
``What surface maximizes entanglement entropy?,''
Phys. Rev. D \textbf{90}, no.8, 085021 (2014)
doi:10.1103/PhysRevD.90.085021
[arXiv:1407.4719 [hep-th]].

\bibitem{BH}
Brown, J.D., Henneaux, M. ``Central charges in the canonical realization of asymptotic symmetries: An example from three dimensional gravity,''.
 Commun.Math. Phys. 104, 207–226 (1986)
 doi:10.1007/BF01211590 .

\end{thebibliography}
\end{document}